\documentclass[11pt,a4paper]{article}

\usepackage{pslatex}
\usepackage{graphicx}
\usepackage{amsmath}
\usepackage{amssymb}
\usepackage{epsfig}
\usepackage{color}
\usepackage[numbers,sort&compress]{natbib}
\usepackage{slashed}

\newcommand{\be}{\begin{equation}}
\newcommand{\ee}{\end{equation}}

\renewcommand{\thefootnote}{\roman{footnote}}
\def\blfootnote{\xdef\@thefnmark{\@footnotetext}}

\title{Elucidating the Vacuum Structure of the Aoki Phase}

\date{\today}
\author{V.~Azcoiti$^\textrm{i}$, G.~Di Carlo$^\textrm{ii}$, E.~Follana$^\textrm{i}$ and A.~Vaquero$^\textrm{iii}$.}

\begin{document}
\footnotetext[1]{Departamento de F\'{\i}sica Te\'orica, Facultad de Ciencias, Universidad de Zaragoza, \\
Cl. Pedro Cerbuna 12, E-50009 Zaragoza (Spain)}

\footnotetext[2]{INFN, Laboratori Nazionali del Gran Sasso, \\
I-67100 Assergi, L'Aquila (Italy)}

\footnotetext[3]{Computation-based Science and Technology Research Center (CaSToRC), The Cyprus Institute, \\
20 Constantinou Kavafi Street, Nicosia 2121, (Cyprus)}

\renewcommand{\thefootnote}{\arabic{footnote}}
\def\blfootnote{\xdef\@thefnmark{\@footnotetext}}

\maketitle

\begin{abstract}
In this paper, we discuss the vacuum structure of QCD with two flavors
of Wilson fermions, inside the Aoki phase. We provide 
numerical evidence, coming from HMC simulations in $4^4$, $6^4$ and $8^4$ 
lattices, supporting a vacuum structure for this model at strong coupling more
complex than the one assumed in the standard wisdom, with new vacua where the
expectation value of $i\bar\psi\gamma_5\psi$ can take non-zero values, and
which can not be connected with the Aoki vacua by parity-flavor symmetry
transformations.
\end{abstract}
	
\section{Introduction}

Two and three colors QCD with unimproved Wilson fermions started to be
simulated in the early 80s \cite{an,hamber}. The complexity of the
phase structure of this model was known from the very beginning, and
indeed the existence of a phase at finite lattice spacing, $a$, with
spontaneous parity and flavor symmetry breaking, was conjectured for
this model by Aoki in the middle 80s \cite{a1,a2}. From that time on,
much work has been done in order to confirm this pattern of
spontaneous symmetry breaking, and to establish a quantitative phase
diagram for lattice QCD with Wilson fermions. References
\cite{abg,creutz,bit,aku,heller,heller2,sharpe,a3,bit2,shasi,kps,gosha,ilg,
ster,golt,ilg3,monos,sharpe2,POS1,POS2,POS3,verba,Splittorff} are an
incomplete list of the work done along these years.

Aoki's conjecture has been supported not only by numerical simulations,
but also by theoretical studies based on the linear sigma model
\cite{creutz} and on applying Wilson chiral perturbation theory
($W\chi PT$) to the continuum effective Lagrangian \cite{sharpe}. The
latter analysis predicts, near the continuum limit, two possible
scenarios, depending on the sign of an unknown low-energy
coefficient. In one scenario, flavor and parity are spontaneously
broken, and there is an Aoki phase with a non-zero value only for the
$i\bar\psi\gamma_5\tau_3\psi$ condensate. In the other one (the
‘‘first-order’’ scenario) there is no spontaneous symmetry breaking.

This standard picture for the Aoki phase was questioned three years
ago by three of us in \cite{monos}, where we conjectured on the
appearance of new vacua in the Aoki phase, which can be characterized
by a non-vanishing vacuum expectation value of the flavor singlet
pseudoscalar condensate $i\bar\psi\gamma_5\psi$, and which can not be
connected with the Aoki vacua by parity-flavor symmetry
transformations. However Sharpe pointed out in \cite{sharpe2} that our
results, based on an analysis of the Probability Distribution Function
(PDF) of fermion bilinears, could still be reconciled with the
standard scenario, and that otherwise we would question also the
validity of the $W\chi PT$ analysis.

For the last few years we have performed an extensive research on the
vacuum structure of the Aoki phase \cite{POS1,POS2,POS3}, in order
to find out if our alternative vacuum structure, derived from the PDF
analysis, was realized or not. The purpose of this paper is to clarify
these issues by reporting the results of our investigations which, as
will be shown along this article, provide evidence of a more complex
vacuum structure in the Aoki phase of two flavor QCD, as conjectured
in \cite{monos}.

The outline of the paper is as follows. Section 2 summarizes the
standard picture on the phase diagram of QCD with two flavors of
unimproved Wilson fermions and our alternative scenario, derives some
interesting expressions to be compared with continuum results, and
analyzes the relevance of a discrete symmetry $P'$, composition of
parity and discrete flavor transformations, which was introduced by
Sharpe and Singleton \cite{sharpe} as an attempt to reconcile the
Vafa-Witten theorem on the impossibility to break spontaneously parity
in a vector-like theory, with the existence of the Aoki phase. Section
3 contains technical data of our HMC simulations of QCD with two
flavors of Wilson fermions, inside and outside the Aoki phase,
performed without external sources and also with a twisted mass term
in the action. The results of our analysis of the PDF of the operator
$Q$, an order parameter for the $P'$ symmetry, are reported in Section
4. This section contains what is, in our opinion, the strongest
indication on the existence of a vacuum structure in the Aoki phase
more complex than the standard one. In Section 5 we show how we were
able to perform a direct measurement of
$\left\langle\left(i\bar\psi\gamma_5\psi\right)^2\right\rangle$ in the
Aoki phase, which produced a non-zero value for this operator in the
$6^4$ lattice, of the same order of magnitude as
$\left\langle\left(i\bar\psi_u\gamma_5\psi_u\right)^2\right\rangle$
and
$\left\langle\left(i\bar\psi\gamma_5\tau_3\psi\right)^2\right\rangle$
(both are non-vanishing in the Aoki phase in the standard picture).
In Section 6 we discuss possible reasons for the discrepancy between
our results and the predictions of $W\chi PT$. Section 7 analyzes the 
new symmetries of the model at $m_0 = -4.0$ as well as the 
behavior of the condensates near the $m_0 = -4.0$ line of the phase diagram,
giving theoretical support to the numerical  results reported in this
article. Finally Section 8 contains our conclusions.

\section{The Aoki phase for the two flavor model}

\begin{figure}[h]
\begin{center}
\resizebox{12 cm}{!}{\includegraphics{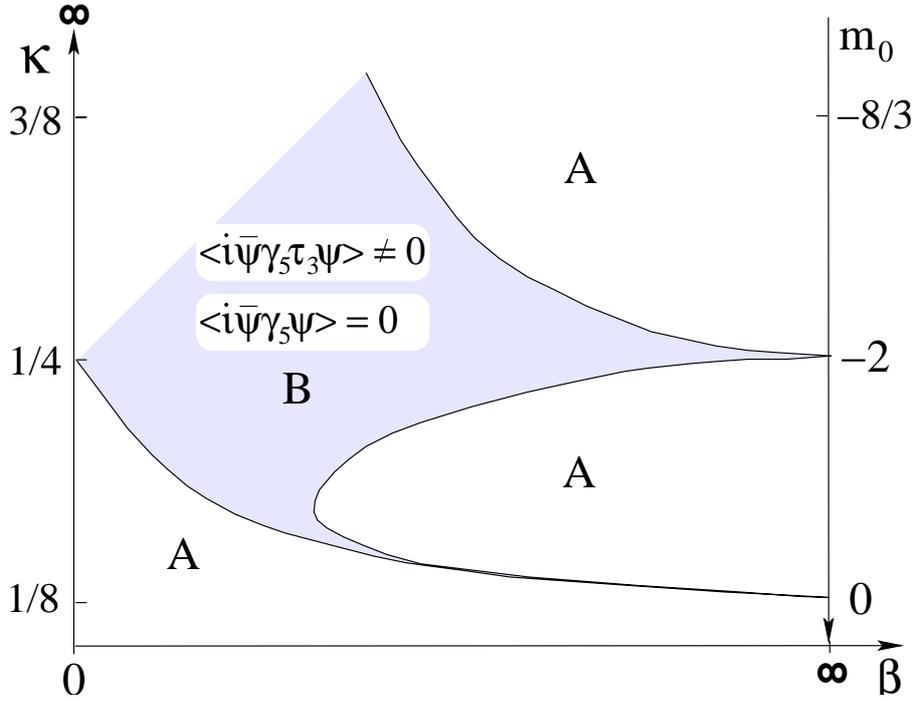}}
\caption{\label{phaseFig}Aoki (B) and physical (A) region in the
  $(\beta,\kappa)$ plane. Adapted from \cite{ilg} by courtesy of the
  authors.}
\end{center}
\end{figure}

The fermionic Euclidean action of QCD with two degenerate Wilson
flavors is

\be
S_F=\bar\psi_u W(\kappa) \psi_u + \bar\psi_d W(\kappa) \psi_d,
\label{acfer}
\ee
where $W(\kappa)$ is the Dirac-Wilson operator, and $\kappa$ is the
hopping parameter, related to the adimensional bare fermion mass $m_0$ 
by $\kappa =1/(8+2m_0)$. The standard wisdom on the phase diagram of
this model in the gauge coupling $\beta, \kappa$ plane is the one
shown in figure \ref{phaseFig}. The two different regions observed in
this phase diagram, A and B, can be characterized as follows: in
region A parity and flavor symmetries are realized in the vacuum,
which is supposed to be unique. The continuum limit should be obtained
by taking the $g^2\rightarrow 0$, $\kappa\rightarrow 1/8$ limit from
within region A, which we will call the QCD region. In region B, on
the contrary, parity and flavor symmetries are spontaneously broken,
there are many degenerate vacua connected by parity-flavor
transformations in this region, and the Gibbs state is therefore made
up from many degenerate vacuum states. In what follows we will call
region B the Aoki region.

\subsection{Standard Wisdom}

The standard order parameters to distinguish the Aoki region from the
QCD region are $i\bar\psi\gamma_5\tau_j\psi$ and
$i\bar\psi\gamma_5\psi$, with $\tau_j$ the three Pauli matrices. For
the standard election, $j=3$, they can be written as a function of the
up and down quark fields as follows

\begin{equation}
\begin{aligned}
i\bar\psi\gamma_5\tau_3\psi &= i\bar\psi_u\gamma_5\psi_u-i\bar\psi_d\gamma_5\psi_d, \\
i\bar\psi\gamma_5\psi &= i\bar\psi_u\gamma_5\psi_u+i\bar\psi_d\gamma_5\psi_d.
\label{orpa}
\end{aligned}
\end{equation}

The standard wisdom for the Aoki phase can be summarized by the
following two equations \cite{a2}

\begin{equation}
\begin{aligned}
\langle i\bar\psi\gamma_5\tau_3\psi \rangle & \neq 0, \\
\langle i\bar\psi\gamma_5\psi \rangle & = 0.
\label{orpaao}
\end{aligned}
\end{equation}

These equations should hold in the vacuum selected when we add a
twisted mass term to the Euclidean Wilson action \eqref{acfer},

\be
S_{m_t} = m_t i\bar\psi\gamma_5\tau_3\psi,
\label{twisted}
\ee 
which breaks both parity and flavor symmetries, and take the vanishing
twisted mass limit once the infinite degrees of freedom limit has been
performed.  The first of the two condensates
$i\bar\psi\gamma_5\tau_3\psi$ breaks both parity and flavor
symmetries.  The non-vanishing vacuum expectation value of this
condensate signals the spontaneous breaking of the $SU(2)$ flavor
symmetry down to $U(1)$, with two Goldstone charged
pions. Notwithstanding parity is spontaneously broken if the first
equation in \eqref{orpaao} holds, the vacuum expectation value of the
flavor diagonal condensate $\langle i\bar\psi\gamma_5\psi\rangle$
vanishes.  Indeed it is also an order parameter for a discrete
symmetry $P'$, composition of parity and discrete flavor rotations,
which is assumed to be realized \cite{sharpe}.

The Aoki phase can then be characterized by the presence of two
massless charged pions and a massive neutral pion which becomes
massless just at the critical line separating this phase from the
physical one. In addition the $\eta$-meson, the mass of which can be
extracted from the long distance behavior of the two-point correlation
function of the flavor singlet pseudoscalar operator
$i\bar\psi\gamma_5\psi$, stays massive even at the critical line as a
consequence of the $U(1)_A$ axial anomaly.

\subsection{Our alternative Scenario}

Three years ago this standard scenario was revised in \cite{monos}
with the help of the probability distribution function (PDF) of the
parity-flavor fermion bilinear order parameters
\cite{pdc}\cite{vw2}. We found that if the spectral density
$\rho_U(\lambda,\kappa)$ of the overlap Hamiltonian $\gamma_5
W(\kappa)$, or Hermitian Dirac-Wilson operator, in a fixed background
gauge field $U$ were not symmetric in $\lambda$ in the thermodynamic
limit for the relevant gauge configurations, Hermiticity of
$i\bar\psi_u\gamma_5\psi_u+i\bar\psi_d\gamma_5\psi_d$ would be
violated at finite $\beta$. Assuming that the Aoki phase ends at
finite $\beta$, this would imply the loss of any physical
interpretation of this phase in terms of particle excitations. If on
the contrary the spectral density $\rho_U(\lambda,\kappa)$ of the
Hermitian Dirac-Wilson operator in a fixed background gauge field $U$
is symmetric in $\lambda$, we argued in \cite{monos} on the appearance
of other phases, which can be characterized by a non-vanishing vacuum
expectation value of
$i\bar\psi_u\gamma_5\psi_u+i\bar\psi_d\gamma_5\psi_d$, and which can
not be connected with the Aoki vacua by parity-flavor symmetry
transformations. We summarize here the main steps of \cite{monos}.

The key point was to study the Fourier transform of the PDF of the
fermion bilinear order parameters, $P(q)$, defined as

\be
P(q) = {\frac{\int [dU] [d\bar\psi d\psi]
\exp \{-S_{YM} +\bar\psi W(\kappa)\psi+ {\frac{iq}{V}} \int d^4x
\>\>  O(x)\}}{\int [dU] [d\bar\psi d\psi] \exp
\{-S_{YM}  +\bar\psi W(\kappa)\psi \}}, }
\label{pdiqferm}
\ee
where $O$ is any fermion bilinear operator

\be
O(x)=\bar\psi(x) \tilde O \psi(x),
\label{bili}
\ee
with $\tilde O$ a matrix with Dirac, color and flavor indices, and $V$ 
is the number of lattice points. Integrating out the fermion fields
one gets

\be
P(q)= {\frac{\int [dU] e^{-S_{YM}} \det(W(\kappa)+{\frac{iq}{V}} \tilde O) }
{\int [dU] e^{-S_{YM}} \det W(\kappa) }, }
\label{pdiqbi}
\ee

\noindent
which can also be expressed as the following mean value

\be P(q)=\left\langle {\frac{\det (W(\kappa)+{\frac{iq}{V}} \tilde
    O)}{\det W(\kappa)}} \right\rangle,
\label{pdiqb}
\ee

\noindent
computed in the effective gauge theory with the integration measure

$$
[dU] e^{-S_{YM}} \det W(\kappa).
$$
The q-derivatives of $P(q)$ give us the moments of the PDF of our
fermion bilinear

\be
{ \frac{d^n{P(q)}}{dq^n}} \bigg\rvert_{q=0} = i^n 
\left\langle \left(\frac{1}{V} \int d^4x\>\>  O(x)\right)^n \right\rangle.
\label{qder}
\ee
For the case we are interested in, and if we call $P_j(q), P_0(q)$ the
PDF of

\begin{equation}
\begin{aligned}
c_j &= i\bar\psi\gamma_5\tau_j\psi,\\
c_0 &= i\bar\psi\gamma_5\psi.
\label{c0c3}
\end{aligned}
\end{equation}
In momentum space, we have

\begin{equation}
\begin{aligned}
P_1(q) = P_2(q) = P_3(q) &= \left\langle \prod_j \left ( - {\frac{q^2}{V^2
\lambda_j^2}} +1 \right ) \right \rangle,\\
P_0(q) &=  \left\langle \prod_j \left ( {\frac{q}{V \lambda_j}} +1 \right )^2
 \right \rangle,
\label{pdfmom}
\end{aligned}
\end{equation}

\noindent
where $\lambda_j$ are the real eigenvalues of the one flavor Hermitian
Dirac-Wilson operator $\bar W(\kappa) = \gamma_5 W(\kappa)$.

Since $c_0, c_j$ are order parameters for the symmetries of the
lattice action and we are not selecting a particular vacuum state,
their first moment will vanish always, independently of the
realization of the symmetries.  The first non-vanishing moment, if the
symmetry is spontaneously broken, will be the second:

\begin{equation}
\begin{aligned}
\left\langle c_0^2 \right\rangle &= 2 \left\langle {\frac{1}{V^2}}
\sum_j {\frac{1}{\lambda_j^2}}\right \rangle  - 4 \left\langle 
\left ( {\frac{1}{V}} \sum_j {\frac{1}{\lambda_j}} \right )^2 \right\rangle,\\
\left\langle c_3^2 \right\rangle &= 2 \left\langle {\frac{1}{V^2}}
\sum_j {\frac{1}{\lambda_j^2}}\right \rangle.
\label{vasp}
\end{aligned}
\end{equation}

In the QCD region flavor symmetry is realized. The PDF of $c_3$ will
be then $\delta(c_3)$ and $\langle c_3^2\rangle = 0$. We get then

$$
\left\langle c_0^2 \right\rangle = - 4 \left\langle 
\left ( {\frac{1}{V}} \sum_j {\frac{1}{\lambda_j}} \right )^2 \right\rangle, 
$$

\noindent
which should vanish since parity is also realized in this
region. Furthermore a negative value of $\langle c_0^2\rangle$ would
violate Hermiticity of $i\bar\psi\gamma_5\psi$.

In the Aoki region \cite{a2} there are vacuum states in which the
condensate $c_3$ \eqref{c0c3} takes a non-vanishing vacuum expectation
value.  This implies that the PDF $P(c_3)$ will not be $\delta(c_3)$
and therefore $\langle c_3^2\rangle$ \eqref{vasp} will not
vanish. Indeed expression \eqref{vasp} for $\langle c_3^2\rangle$
seems to be consistent with the Banks and Casher formula \cite{BC}
which relates the spectral density of the Hermitian Dirac-Wilson
operator at the origin with the vacuum expectation value of $c_3$
\cite{sharpe}.

If, on the other hand, $\langle i\bar\psi\gamma_5\psi\rangle=0$ in one
of the Aoki vacua, as conjectured in \cite{a2}, $\langle
i\bar\psi\gamma_5\psi\rangle=0$ in all the other vacua that are
connected with the standard Aoki vacuum by a parity-flavor
transformation, since $i\bar\psi\gamma_5\psi$ is invariant under
flavor transformations and changes sign under parity. Therefore if we
assume that these are all the degenerate vacua, we should conclude
that $P(c_0)= \delta(c_0)$ and $\langle c_0^n\rangle=0$, which would
imply an infinite tower of non-trivial relations, one for each even
moment of the PDF. We write here the simplest one,

\be
\left\langle {\frac{1}{V^2}}
\sum_j {\frac{1}{\lambda_j^2}}\right \rangle  = 2 \left\langle
\left ( \frac{1}{V} \sum_j {\frac{1}{\lambda_j}} \right )^2 \right\rangle
\neq 0.
\label{rela}
\ee

A possible scenario which was discussed in \cite{monos} is the one
that emerges if we assume a symmetric distribution for the eigenvalues
of the Hermitian Wilson operator. As discussed in \cite{monos} this
assumption is necessary in order to preserve Hermiticity. If we take
this assumption in the most naive way we get

\be
\langle c_3^2 \rangle - \langle c_0^2 \rangle = 
4 \left\langle \left ( \frac{1}{V} \sum_j {\frac{1}{\lambda_j}} \right )^2
\right \rangle = 0,
\label{diffe}
\ee
and therefore we should conclude that new vacua characterized by a
non-vanishing vacuum expectation value of the singlet flavor pseudoscalar
operator $i\bar\psi\gamma_5\psi$ must exist besides the Aoki vacua.
Nevertheless as Sharpe noticed in \cite{sharpe2}, sub-leading
contributions to the spectral density may affect equation
\eqref{diffe} in the $\epsilon$-regime in such a way that, not only
$\langle c_0^2 \rangle$, but every even moment of
$i\bar\psi\gamma_5\psi$ would vanish, restoring the standard Aoki
picture.

As we will show in this article equation \eqref{diffe} turns out not
to be realistic in the Aoki phase, as follows from numerical
simulations.  However the thesis of Sharpe, although possible and
inspired in the absence of our conjectured new vacua in the chiral
effective Lagrangian approach, would enforce as discussed before an
infinite series of \emph{sum rules}, similar to those found by
Leutwyler and Smilga in the continuum \cite{Leut}. The main purpose of
this paper is to clarify these issues.

\subsection{Some interesting relations}

As previously discussed flavor and parity symmetries should be
realized in the vacuum of the physical phase. This phase can therefore
be characterized, in what concerns its low energy spectrum, by the
existence of three degenerate massive pions, which become massless at
the critical line, and the $\eta$-meson, which is massive even at the
critical line because of the axial anomaly. When we cross the critical
line and enter into the Aoki phase, the neutral pion becomes massive
whereas the charged pions are massless since they are the two
Goldstone bosons associated to the spontaneous breaking of the $SU(2)$
flavor group to a $U(1)$ group.

The susceptibilities of the neutral pion $\chi_{\pi^0}$ and the
$\eta$-meson $\chi_\eta$ in the physical phase are the corresponding
integrated two-point correlation functions

\begin{align}
\chi_{\pi^0} =& \sum_x \left\langle i\bar\psi\gamma_5\tau_3\psi \left(x\right)
i\bar\psi\gamma_5\tau_3\psi \left(0\right)\right\rangle &=& \frac{2}{V}\left\langle
\sum_{i=1}^V\frac{1}{\lambda_i^2}\right\rangle\label{SusPi},\\
\chi_{\eta} =& \sum_x \left\langle i\bar\psi\gamma_5\psi \left(x\right)
i\bar\psi\gamma_5\psi \left(0\right)\right\rangle\textrm{\footnotemark} &=& 
\frac{2}{V}\left\langle \sum_{i=1}^V\frac{1}{\lambda_i^2}\right\rangle - \frac{4}{V}\left\langle
\left[\sum_{i=1}^V\frac{1}{\lambda_i}\right]^2 \right\rangle\label{SusEta}.
\end{align}
\footnotetext{There are no disconnected contributions in this phase.}

The rightmost hand sides of these equations are just the second
moments of the PDF of $i\bar\psi\gamma_5\tau_3\psi$ and
$i\bar\psi\gamma_5\psi$ (see equations \eqref{vasp}) multiplied by the
corresponding volume factor.  Therefore equations \eqref{SusPi} and
\eqref{SusEta} give us the following relation between $\chi_{\eta}$,
$\chi_{\pi^0}$ and the trace of the inverse one flavor Hermitian
Dirac--Wilson operator:

\be
\chi_{\eta} = \chi_{\pi^0} - {4\over V} \left\langle Tr^2 
\left(\gamma_5 W^{-1}\left(\kappa\right)\right)\right\rangle,
\label{sustop}
\ee
to be compared with the well known relation between the eta, pion and 
topological susceptibilities

\be
\chi_{\eta} = \chi_{\pi^0} + 4 {\chi_T\over m^2},
\label{sustopgw}
\ee
which holds in the continuum and also in the Ginsparg-Wilson
regularization at finite lattice spacing.

The first interesting observation that follows from equations
\eqref{sustop} and \eqref{sustopgw} is the suggestive relation
between the topological susceptibility $\chi_T$ and the trace of the
inverse Hermitian Dirac--Wilson operator,

\be
{\chi_T\over m^2} = - {1\over V} \left\langle Tr^2
\left(\gamma_5 W^{-1}\left(\kappa\right)\right)\right\rangle,
\label{relation}
\ee
which should hold near the continuum limit. This suggests also the
following relation between the trace of the inverse Hermitian
Dirac--Wilson operator, quark mass $m$, and the density of topological
charge $q = Q/V$ \cite{smit, victor}

\be
{q\over m} = {1\over V} Tr
\left(\gamma_5 W^{-1}\left(\kappa\right)\right),
\label{relation2}
\ee

A second interesting observation that follows from \eqref{sustop}
concerns the behavior of

\be
{4\over V} \left\langle Tr^2
\left(\gamma_5 W^{-1}\left(\kappa\right)\right)\right\rangle = 
\frac{4}{V}\left\langle\left[\sum_{i=1}^V\frac{1}{\lambda_i}\right]^2
\right\rangle
\label{traza}
\ee 
when we approach the critical line from the physical phase. Indeed in
the physical phase this quantity should be finite because the pion and
eta susceptibilities are finite. However when approaching the critical
line it should diverge in order to compensate the divergence of the
pion susceptibility keeping $\chi_\eta$ finite in \eqref{sustop}, in
deep similarity with the divergence of the topological susceptibility
divided by the square quark mass in the continuum and Ginsparg--Wilson
regularization in the chiral limit. The origin of this divergence lies
in the accumulation of eigenvalues of the Hermitian Dirac--Wilson
operator near the origin.

The divergence of \eqref{traza} at the critical line is slower than
$V$ in such a way as to keep a parity-flavor symmetric vacuum, as
corresponds to a second order phase transition. On the other hand if
the standard scenario for the Aoki phase is realized, the higher
accumulation of eigenvalues of the hermitian Dirac-Wilson operator at
the origin in this phase should be enough to give a finite
contribution to $\left\langle c^2_i\right\rangle$ (second equation in
\eqref{vasp}).  Indeed if the Aoki vacuum selected with a twisted mass
term plus those obtained by parity-flavor symmetry transformations are
the only degenerate vacua in this phase, the following relation should
hold \cite{sharpe2,pdc}

\be
\left\langle c_3^2 \right\rangle = {\Sigma^2\over 3},
\label{standard}
\ee
where $\Sigma$ is the mean value of $c_3$ in the vacuum selected by the 
twisted mass term. In Section 5 we will report the results of a check of
equation \eqref{standard}.

The standard scenario for the Aoki phase requires also that the
singlet flavor pseudoscalar operator $i\bar\psi\gamma_5\psi$ takes a
vanishing vacuum expectation value. Taking into account the first
equation in \eqref{vasp}, this requirement implies that the l.h.s. of eq. 
\eqref{traza}, ${4\over V} \left\langle Tr^2 \left(\gamma_5
W^{-1}\left(\kappa\right)\right)\right\rangle$, should diverge as $V$,
to compensate the first contribution to $\left\langle
c^2_0 \right\rangle$ in \eqref{vasp}.

\subsection{Spontaneous Symmetry Breaking of P'?}

Years ago Sharpe and Singleton suggested \cite{sharpe} that a symmetry
$P'$, composition of parity and flavor transformations, associated to
a discrete subgroup of the parity-flavor continuous group, and acting
on pure gluonic operators as parity, could still be realized in the
Aoki phase, notwithstanding parity and flavor are spontaneously broken
in this phase. Their motivation for such a proposal was the attempt to
reconcile the Vafa--Witten theorem \cite{witten} on the impossibility
to break spontaneously parity in a vector-like theory, with the
existence of the Aoki phase. Their point was that Vafa--Witten theorem,
although it does not apply to fermionic order parameters, could still
apply to pure gluonic operators. Then if the realization of $P'$ does
not require a vanishing Aoki condensate, and $P'$ acts on pure gluonic
operators as parity, the existence of the Aoki phase besides the
realization of $P'$ in the vacuum would not be in conflict with the
Vafa--Witten theorem for pure gluonic operators. Furthermore the
realization of $P'$ would be quite useful since it would give a simple
explanation for the tower of sum rules \eqref{rela} required in the
standard scenario. Indeed the flavor singlet pseudoscalar condensate
$i\bar\psi\gamma_5\psi$ is an order parameter for $P'$.

Although the main motivation for the introduction of the $P'$
symmetry, the realization of Vafa--Witten theorem for pure gluonic
operators, has become less relevant on the light of later works on
this subject \cite{monos2}-\cite{monos3}, the implications on the
standard scenario for the Aoki phase of the realization of this
symmetry are still relevant.

A possible election for $P'$, the realization of which would not be in
conflict with non-vanishing condensates $\left\langle
i\bar\psi\gamma_5 \tau_j\psi\right\rangle$ for $j = 1, 2, 3$ is the
composition of parity with the $Z_4$ subgroup of the $SU(2)$ flavor
group generated by

$$
{i\over{\sqrt 3}} \left( \tau_1 + \tau_2 + \tau_3 \right).
$$

In this context let us consider the operator

\be
Q = {1\over V} Tr
\left(\gamma_5 W^{-1}\left(\kappa\right)\right) = 
\frac{1}{V} \sum_{j=1}^V\frac{1}{\lambda_j},
\label{traza2}
\ee 
which is a non-local order parameter for $P'$. This operator, although
non-local, is not singular since exact zero modes have vanishing
integration measure in the Wilson formulation. We will assume in what
follows that $Q$ is an intensive operator. Notice that any local
operator should be intensive but any intensive operator is not
necessarily local. Our assumption is based in the following argument:
$\left\langle Q\right\rangle = {1\over 2V}
\left\langle\sum_x\bar\psi(x)\gamma_5\psi(x)\right\rangle$, 
$\bar\psi(x)\gamma_5\psi(x)$ being a local fermionic operator. Our
prejudices tell us that ${1\over 2V}\sum_x\bar\psi(x)\gamma_5\psi(x)$, 
as any intensive operator, should be self-averaging. In other words,
if the system size is very large, a single thermalized gauge
configuration should be enough to get ${1\over 2V}
\left\langle\sum_x\bar\psi(x)\gamma_5\psi(x)\right\rangle$, and this
seems indeed to be the case in the physical phase, where the PDF of
$Q$ approaches a delta, as shown in Section 4. Furthermore equations
\eqref{sustop} and \eqref{sustopgw} strongly suggest that, near the
continuum limit, $Q$ would be essentially the density of topological
charge, which indeed is an intensive operator.

Hence were the $P'$ symmetry realized in the vacuum, the PDF of
\eqref{traza2} would be, according to our assumption, a $\delta(x)$
distribution. But the second moment of this PDF,

$$
\frac{1}{V^2}\left\langle\left[\sum_{i=1}^V\frac{1}{\lambda_i}\right]^2
\right\rangle,
$$ 
needs to be finite in the standard scenario for the Aoki phase in
order to realize the first sum rule \eqref{rela}. Therefore the
standard scenario requires the spontaneous symmetry breaking of $P'$,
but then there are no symmetry reasons to justify this tower of sum
rules. In Section 4 we will report numerical results for the PDF of
\eqref{traza2}, both in the Aoki and in the physical phase.

\section{The simulations}
\label{simulations}

In order to determine the phase structure and the properties of the
vacuum, we need to measure the $PDFs$ of the parity-flavor order
parameters described in the previous sections, including the $Q$
operator \eqref{traza2}. So we decided to carry out HMC simulations
of QCD with two flavors of Wilson fermions, inside and outside the
Aoki phase, and mainly without external sources, although some
simulations with a twisted mass term in the action have also been
performed.

We remove the external sources because of several facts:

\begin{enumerate}
\item The PDF formalism requires the removal of any external sources
  in the action. We just perform one long simulation in the
  Gibbs state.
\item The addition of a twisted mass external source, as has been done
  in past simulations of the Aoki phase, automatically selects the
  vacuum where the standard properties of the Aoki phase are verified
  \cite{monos}. This point could explain why no one ever saw a new
  Aoki phase like the one we are proposing, since all the past
  simulations performed with two flavors of Wilson fermions inside the
  Aoki phase were done with an external twisted mass term.
\item We could try to select one of these hypothetical new vacua by
  adding an external source proportional to $i\bar\psi\gamma_5\psi$,
  but this introduces a severe sign problem in the simulations.
\end{enumerate}

The Aoki phase without external sources is very hard to
simulate. Inside the Aoki phase there are exactly massless pions and
quasi-zero modes, rendering the condition number of the Wilson Dirac
operator quite high. A standard solver will not be enough to invert
the Dirac operator in a HMC simulation. Fortunately, in recent years a
number of efficient solvers have appeared, and by using a DD--HMC
\cite{luescher} we could invert the Dirac operator at a reasonable
speed with the resources available to us, i.e., the clusters of the
department of theoretical physics of the University of Zaragoza,
comprising 160 cores and 280 Gb of available memory, interconnected by
a gigabit network. The simulations were parallelized for 4, 8 or 12
cores using openMP, and we did several simultaneous runs. The
iteration count of the solver ranged from a handful outside the Aoki
phase, to a few hundred inside the Aoki phase in the largest volume
$8^4$ and without external sources. For volumes higher than $8^4$, the
inversion time became prohibitive for our computing resources and our
time-frame, and we should look for new ways of simulating the Aoki
phase.

Moreover, there is an additional problem with these simulations: since
quasi-zero modes appear inside the Aoki phase, the eigenvalues
sometimes attempt to cross the origin, and they would do so, were it
not for the fact that the crossing of eigenvalues is forbidden by the
HMC dynamics: the potential energy in the molecular dynamics step
diverges at the origin, and there is an infinite repulsion that
prevents the eigenvalues from crossing.

In order to solve this problem we classified our simulations according
to its sector number, i.e., the number of 'crossed' eigenvalues of the
Hermitian Dirac--Wilson operator they had: beginning from a completely
symmetric state (same number of positive and negative eigenvalues),
the number of the eigenvalues that should cross to reach the desired
state; and we performed simulations with a different number of crossed
eigenvalues for each volume. Now this does not completely solve the
problem, since we still don't know the weight of each sector within
the partition function. The only way to simulate dynamically all the
sectors is to add a twisted mass term, but as explained above we are
primarily interested in the results without external
sources. Fortunately the weights of the different sectors evolve very
slowly with the twisted mass, and we can then extrapolate the data to 
vanishing twisted mass.

Indeed in Table \ref{Weights} we report the weights of the different
sectors as a function of the volume and twisted mass $m_t$. \emph{Qch}
stands for quenched, whereas the numbers marked as MFA come from an
$MFA$ (Microcanonical Fermion Average) inspired approach \cite{MFA},
and were obtained by diagonalizing $4\times10^6$ gauge configurations. As
can be seen in this table the weights show a very mild
$m_t$-dependence, although the $4^4$ results at $m_t = 0$ obtained
from $MFA$ simulations are at two standard deviations from the
extrapolated results. It is not clear to us if this discrepancy has a
statistical origin or reflects a discontinuity of the weights of
different sectors at $m_t = 0$. However if the latter were the actual
case, the standard picture for the Aoki phase could not be realized
since it requires continuity in the sample of relevant gauge
configurations at $m_t = 0$. Hence we will assume continuity of the
weights, and in the following sections we will use Table \ref{Weights}
to reconstruct the PDF for the interesting bilinears by summing the
weighted PDFs of all sectors for a given volume. The details of the
runs can be checked in Table \ref{Sims}.

\begin{table}[h!]
\begin{center}
\footnotesize
\begin{tabular}{|c|c|c|c|c|c|}
\hline Vol. & $m_t$ & Sector 0 & Sector 1 & Sector 2 & Sector 3+ \\
\hline
$4^4$ & MFA & $80\% \pm6\%$ & $20\% \pm6\%$ & $0\%$ & $0\%$ \\
$4^4$ & $0.01$ & $68.1\%\pm1.0\%$ & $31.6\%\pm1.0\%$ & $0.24\%\pm0.08\%$ & $0\%$ \\
$4^4$ & $0.05$ & $70.4\%\pm0.3\%$ & $29.4\%\pm0.3\%$ & $0.18\%\pm0.07\%$ & $0\%$ \\
$4^4$ & $0.10$ & $69.3\%\pm0.9\%$ & $30.4\%\pm0.9\%$ & $0.28\%\pm0.04\%$ & $0\%$ \\
$4^4$ & $0.30$ & $70.3\%\pm0.7\%$ & $29.6\%\pm0.6\%$ & $0.14\%\pm0.05\%$ & $0\%$ \\
$4^4$ & $0.50$ & $70.0\%\pm0.5\%$ & $29.8\%\pm0.5\%$ & $0.20\%\pm0.06\%$ & $0\%$ \\
$4^4$ & Qch    & $71.4\%\pm0.3\%$ & $28.5\%\pm0.2\%$ & $0.07\%\pm0.02\%$ & $0\%$ \\
$6^4$ & $0.01$ & $51.8\%\pm0.6\%$ & $44.7\%\pm0.7\%$ & $3.3\%\pm0.3\%$ & $0\%$ \\
$6^4$ & $0.05$ & $49.9\%\pm0.8\%$ & $45.7\%\pm0.4\%$ & $4.3\%\pm0.5\%$ & $0.08\%\pm0.02\%$ \\
$6^4$ & $0.10$ & $51.4\%\pm0.6\%$ & $44.7\%\pm0.6\%$ & $3.86\%\pm0.07\%$ & $0.06\%\pm0.02\%$ \\
$6^4$ & $0.30$ & $52.7\%\pm1.2\%$ & $44.3\%\pm1.0\%$ & $2.96\%\pm0.20\%$ & $0.06\%\pm0.06\%$ \\
$6^4$ & $0.50$ & $53.1\%\pm0.6\%$ & $44.0\%\pm0.5\%$ & $2.82\%\pm0.18\%$ & $0.06\%\pm0.02\%$ \\
$6^4$ & Qch    & $55.0\%\pm1.6\%$ & $43.0\%\pm1.9\%$ & $2.0\%\pm0.4\%$ & $0\%$ \\
$8^4$ & $0.05$ & $36.0\%\pm0.6\%$ & $48.2\%\pm0.7\%$ & $13.9\%\pm0.4\%$ & $2.06\%\pm0.20\%$ \\
$8^4$ & $0.10$ & $36.1\%\pm0.6\%$ & $47.0\%\pm0.4\%$ & $14.6\%\pm0.3\%$ & $2.22\%\pm0.28\%$ \\
$8^4$ & $0.30$ & $38.2\%\pm0.4\%$ & $46.8\%\pm0.4\%$ & $13.33\%\pm0.12\%$ & $1.6\%\pm0.4\%$ \\
$8^4$ & $0.50$ & $39.0\%\pm0.3\%$ & $48.5\%\pm0.3\%$ & $11.33\%\pm0.25\%$ & $1.18\%\pm0.11\%$ \\
\hline
\end{tabular}
\caption{Weights of the different sectors as a function of the volume
  and $m_t$.\label{Weights}}
\end{center}
\end{table}
\normalsize

\begin{table}[h!]
\begin{center}
\begin{tabular}{|c|c|c|c|c|c|c|c|c|}
\hline
Vol.  & $\beta$ & $\kappa$ & $m_t$ &    Sector   &  Confs  &    Acc ($\%$)   &        HMC Step        & $l_{Traj}$ \\
\hline
$4^4$ & $2.00$  & $0.25$  & $0.00$ &     $0$     & $5000$  & $99\%$ $(72\%)$ & 3.125e-03 & 0.1 \\
$4^4$ & $2.00$  & $0.25$  & $0.00$ &     $1$     & $2473$  & $99\%$ $(73\%)$ &   2.5e-03 & 0.1 \\
$6^4$ & $2.00$  & $0.25$  & $0.00$ &     $0$     & $676$   & $92\%$ $(60\%)$ &  6.25e-04 & 0.04 \\
$6^4$ & $2.00$  & $0.25$  & $0.00$ &     $1$     & $2000$  & $88\%$ $(56\%)$ &  6.25e-04 & 0.04 \\
$4^4$ & $4.00$  & $0.18$  & $0.00$ & $All^\star$ & $10000$ &      $92\%$     &   0.1    & 1.0 \\
$6^4$ & $4.00$  & $0.18$  & $0.00$ & $All^\star$ & $5000$  &      $84\%$     &   0.1    & 1.0 \\
$8^4$ & $4.00$  & $0.18$  & $0.00$ & $All^\star$ & $5000$  &      $74\%$     &   0.1    & 1.0 \\
$4^4$ & $2.00$  & $0.25$  & $0.01$ &    $All$    & $5001$  &      $83\%$     &   0.0133 & 0.2 \\
$4^4$ & $2.00$  & $0.25$  & $0.05$ &    $All$    & $5001$  &      $86\%$     &   0.05   & 0.4 \\
$4^4$ & $2.00$  & $0.25$  & $0.10$ &    $All$    & $5664$  &      $92\%$     &   0.025  & 0.5 \\
$4^4$ & $2.00$  & $0.25$  & $0.30$ &    $All$    & $4999$  &      $96\%$     &   0.1    & 0.5 \\
$4^4$ & $2.00$  & $0.25$  & $0.50$ &    $All$    & $10000$ &      $99\%$     &   0.025  & 0.5 \\
$6^4$ & $2.00$  & $0.25$  & $0.01$ &    $All$    & $2520$  & $99\%$ $(89\%)$ &   0.01   & 0.4 \\
$6^4$ & $2.00$  & $0.25$  & $0.05$ &    $All$    & $5000$  &      $70\%$     &   0.05   & 0.4 \\
$6^4$ & $2.00$  & $0.25$  & $0.10$ &    $All$    & $5000$  &      $95\%$     &   0.05   & 0.4 \\
$6^4$ & $2.00$  & $0.25$  & $0.30$ &    $All$    & $5001$  &      $90\%$     &   0.1    & 0.5 \\
$6^4$ & $2.00$  & $0.25$  & $0.50$ &    $All$    & $5001$  &      $99\%$     &   0.05   & 0.4 \\
$8^4$ & $2.00$  & $0.25$  & $0.05$ &    $All$    & $5001$  &      $77\%$     &   0.04   & 0.2 \\
$8^4$ & $2.00$  & $0.25$  & $0.10$ &    $All$    & $5001$  &      $84\%$     &   0.05   & 0.5 \\
$8^4$ & $2.00$  & $0.25$  & $0.30$ &    $All$    & $5001$  &      $96\%$     &   0.05   & 0.4 \\
$8^4$ & $2.00$  & $0.25$  & $0.50$ &    $All$    & $10000$ &      $98\%$     &   0.05   & 0.5 \\
\hline
\end{tabular}
\caption{Details of the simulations that generated our
  configurations. The word \emph{Sector} refers to the eigenvalue
  sector where the simulation was performed, as explained above. The
  $^\star$ means that only sector 0 contributes in this case, so all
  sectors were taken into account. Also, \emph{Confs} refers to the
  number of configurations saved (we saved one configuration for each
  two generated, so as to reduce the autocorrelations), \emph{Acc} is
  the acceptance ratio of the simulation (the value in parenthesis
  marks   the acceptance without the replay trick), \emph{HMC Step} is
  the  molecular dynamics time-step used (during replays, the step was
  halved) and $l_{Traj}$ is the trajectory length used. \label{Sims}}
\end{center}
\end{table}
\normalsize The high variability of the acceptance ratio is due to the
hard task of fine-tuning the solver parameters and the trajectory
length to work properly for each case. As the table shows, certain
simulations required a very small time-step for the HMC to work; in
those simulations the eigenvalues were trying to cross the origin,
generating high forces in the molecular dynamics, and thence creating
a high rejection rate, so this small time-step was strictly necessary
to obtain reasonable data. To enhance acceptance, we introduced the
\emph{replay trick} in the hardest calculations, i.e., those inside
the Aoki phase and without a twisted mass term. During replays, the
stepsize was halved whereas the trajectory length was kept constant.

In order to calculate the PDF's we diagonalized the Hermitian
Dirac--Wilson operator on all configurations generated and found the
eigenvalues. The diagonalizations were carried out by using a
parallelized Lanczos algorithm, combined with a Sturm bisection. The
algorithm was checked heavily against the LAPACK library before
starting production to ensure that our results were correct, but our
algorithm was much faster than the LAPACK standard diagonalization
procedure for Hermitian matrices, and used a small fraction of the
memory, because in our algorithm the fermionic matrix was generated
on-the-fly. The diagonalization times ranged from around a few seconds
for the $4^4$ to a few hours in the case of the $8^4$ on our 12 core
machines.

\section{Probability Distribution Function of $Q$}

\begin{figure}[h!]
\begin{center}
\resizebox{13 cm}{!}{\includegraphics[angle=270]{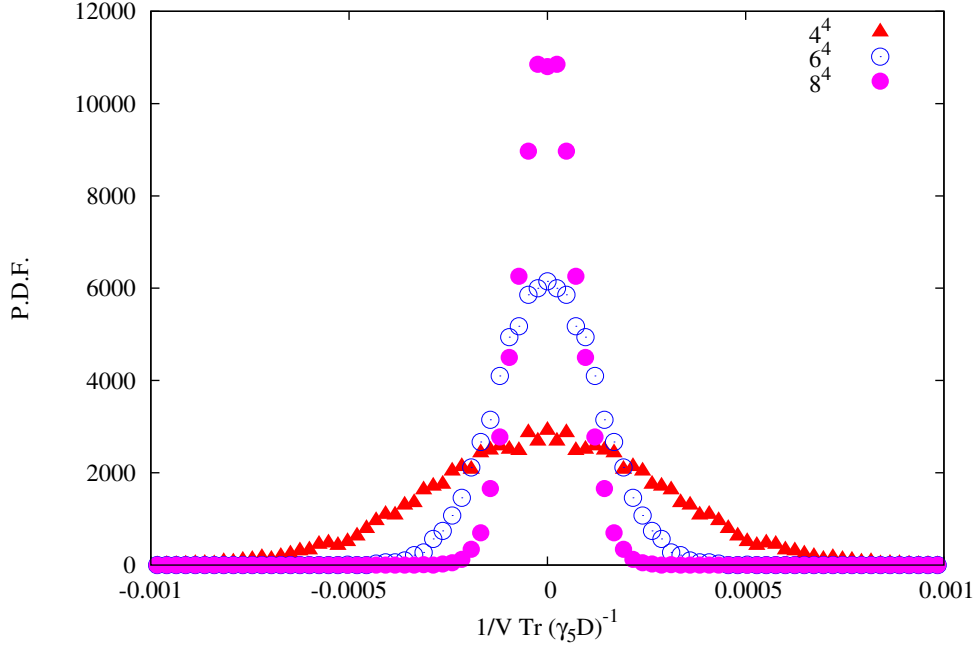}}
\caption{PDF of operator $Q$ outside the Aoki phase for three
  different volumes.\label{OutAoki}}
\end{center}
\end{figure}

First of all, we checked the behavior of the PDF of the operator $Q$
outside the Aoki phase (point at $\beta = 4.0$ and $\kappa = 0.18$) in
the three volumes considered ($4^4$, $6^4$ and $8^4$). As figure
\ref{OutAoki} clearly indicates, the operator $Q$ behaves as an
intensive operator, as expected, and its PDF tends to a Dirac delta at
the origin as the volume increases, showing that both parity and $P'$
symmetries are realized in the vacuum of the physical phase.

\begin{figure}[h!]
\begin{center}
\resizebox{13 cm}{!}{\includegraphics[angle=270]{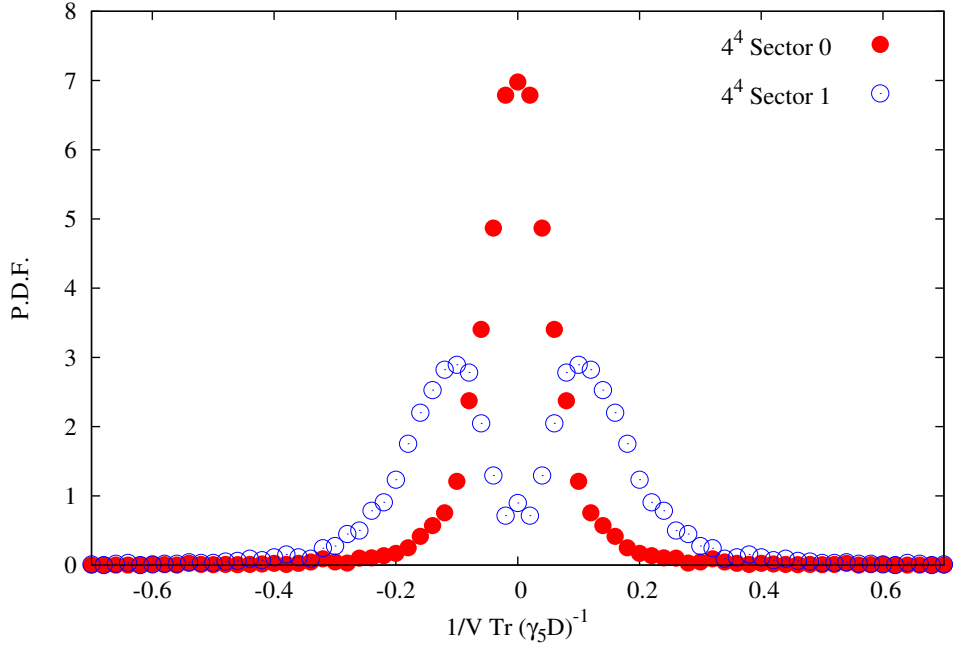}}
\caption{PDF of operator $Q$ inside the Aoki phase for the two sectors
  considered in the $4^4$ case. The sector 1 clearly shows a double
  peaked behavior\label{aokiPlot4}}
\end{center}
\end{figure}

\begin{figure}[h!]
\begin{center}
\resizebox{13 cm}{!}{\includegraphics[angle=270]{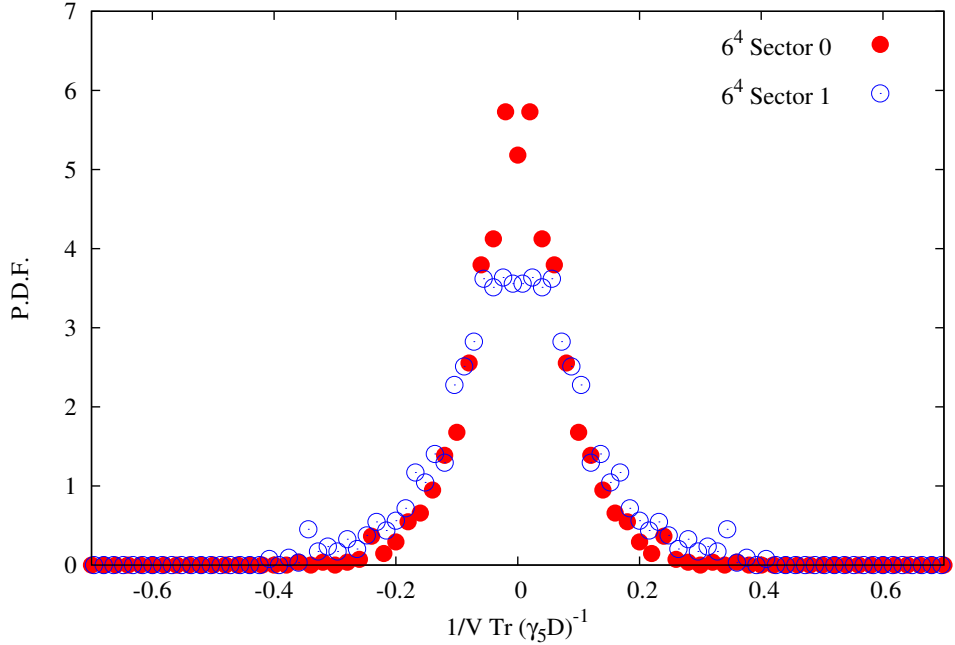}}
\caption{PDF of operator $Q$ inside the Aoki phase for the two sectors
  considered, now in the $6^4$ volume. The double peak of sector 1
  becomes narrower, but we should expect wider contributions from
  higher sectors.\label{aokiPlot6}}
\end{center}
\end{figure}

\begin{figure}[h!]
\begin{center}
\resizebox{13 cm}{!}{\includegraphics[angle=270]{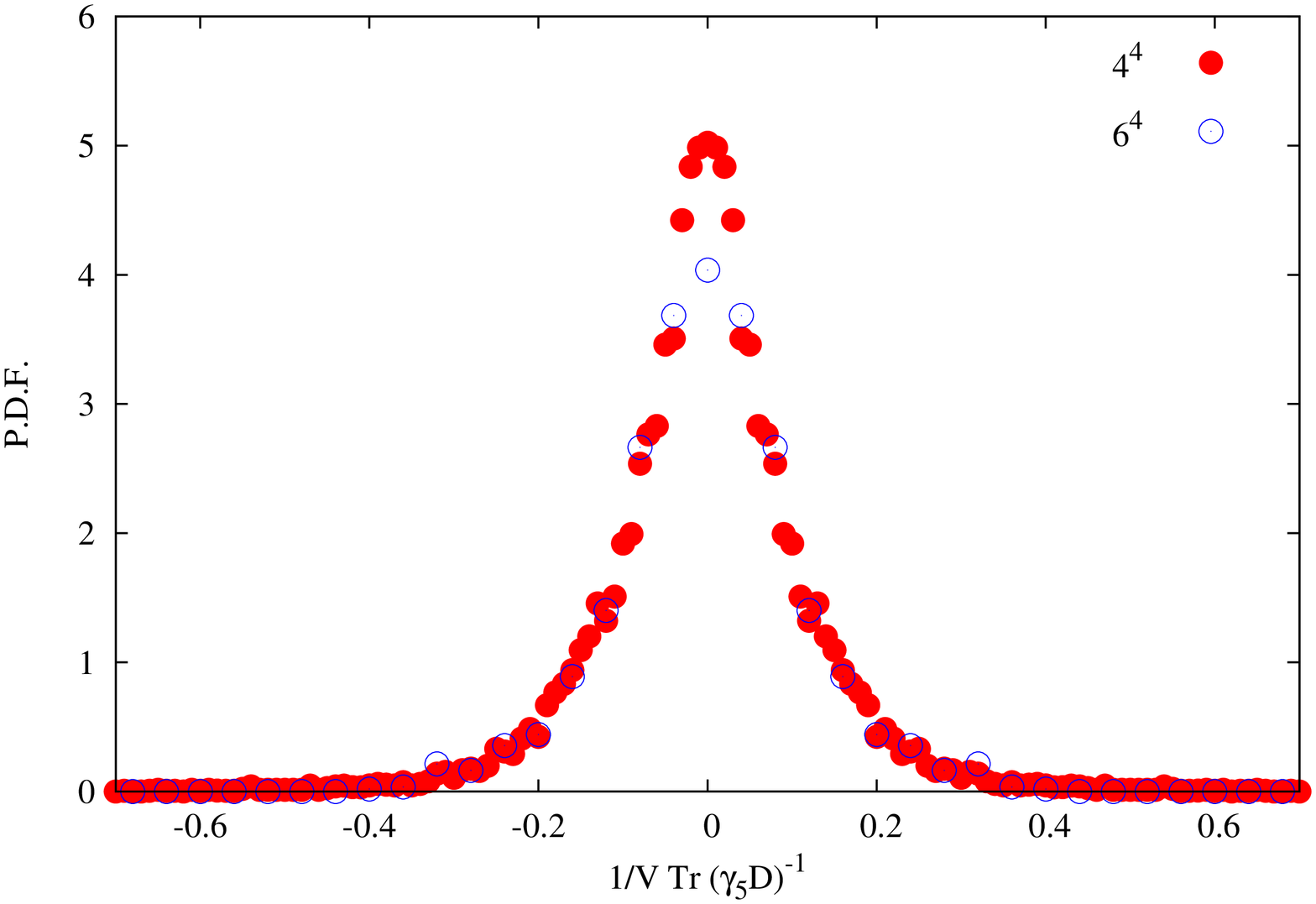}}
\caption{PDF of operator $Q$ inside the Aoki phase for two different
  volumes. The differences in the area can be explained by realizing
  that sector 2 is missing in the $6^4$ volume, which should account
  for a 5\% of the area.\label{aokiPlot}}
\end{center}
\end{figure}

Let's see what happens inside the Aoki phase ($\beta = 2.0$ and
$\kappa = 0.25$, and for this case we will use data from the $4^4$ and
$6^4$ volumes (the $8^4$ inside the Aoki phase was very expensive for
us). We see in figure \ref{aokiPlot4} and figure \ref{aokiPlot6} how
the PDF behaves quite differently depending on the volume and on the
sector. However, when we compute the final result taking into account
the weight of each sector\footnote{The weights are computed by
  extrapolation in $m_t$, assuming continuity in $m_t$, as discussed
  in Section \ref{simulations}.}, we see in figure \ref{aokiPlot} how
all these different plots converge to a single peak of constant width,
in remarkable contrast with the results in the physical phase reported
in figure \ref{OutAoki}. Therefore we expect, according to our
discussion in Section 2.4, that not only parity but also $P'$ will be
spontaneously broken inside the Aoki phase. We should notice that this
result needs to hold if the standard picture of the Aoki phase is
correct, since otherwise $\langle c^2_0 \rangle = \langle c^2_3
\rangle\ne 0$ in this phase. 

Of course we are far from the thermodynamic limit, and one might argue
that, as stated in Table \ref{Weights}, the contribution of sector 2
to the $6^4$ volume is important enough to be taken into account. This
sector was not simulated because it was extremely expensive from the
numerical point of view, and since its weight is less than $5\%$, we
don't expect any changes to be relevant to the final result.

\begin{figure}[h!]
\begin{center}
\resizebox{13 cm}{!}{\includegraphics[angle=270]{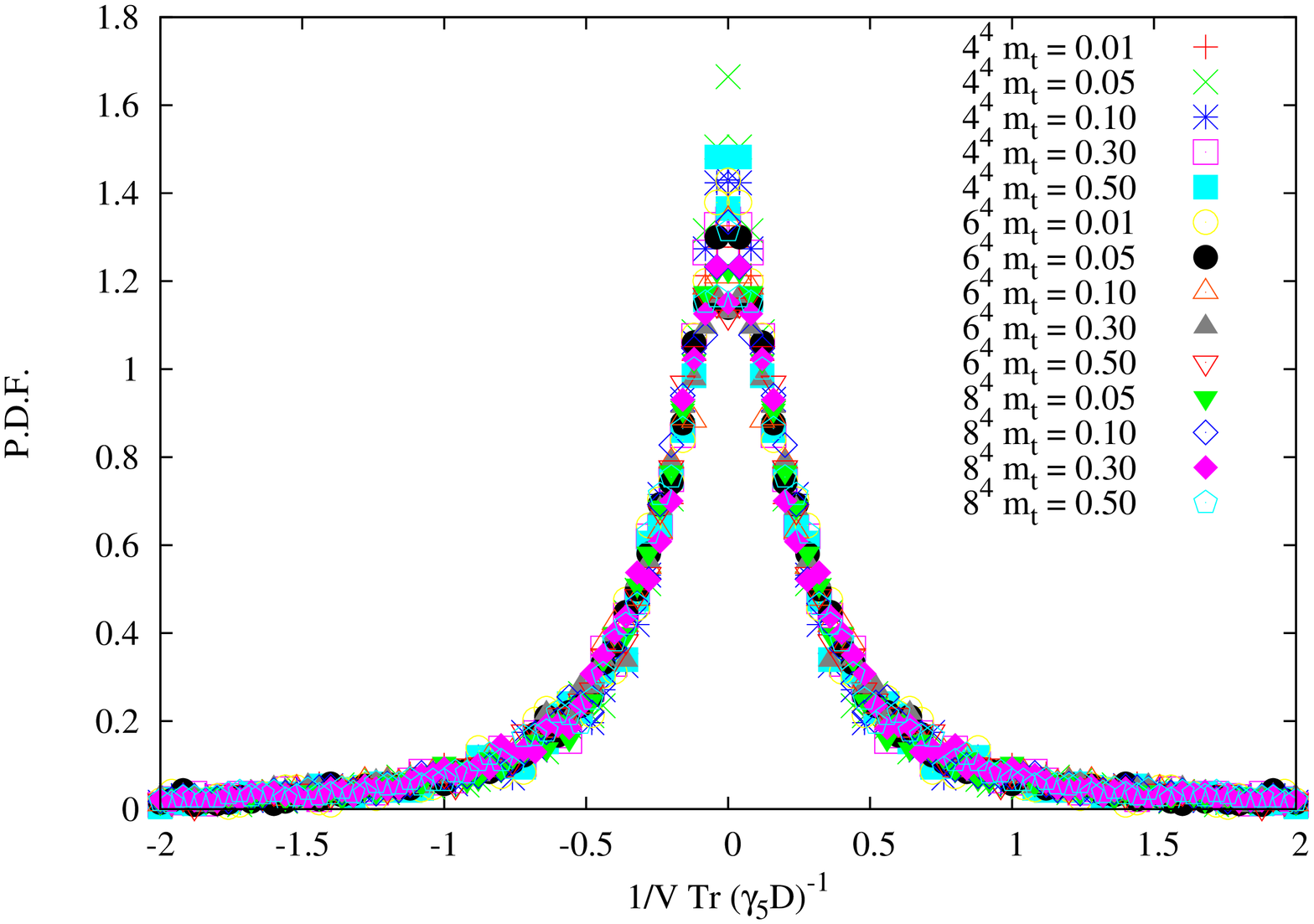}}
\caption{PDF of operator $Q$ inside the Aoki phase for configurations
  generated with a twisted mass source.\label{tmPlot}}
\end{center}
\end{figure}

Another indication that there are new vacua inside the Aoki phase not
considered in the standard picture, comes from measuring the PDF of
the operator $Q$ on configurations dynamically generated with several
values of an external twisted mass term, $m_t
i\bar\psi\gamma_5\tau_3\psi$, but keeping $m_t=0$ in the definition of
$Q$. Since the addition of an external source selects a standard Aoki
vacuum, we expect that all the contributions to the PDF coming from
the other vacua will be removed. As we can see in figure \ref{tmPlot},
the PDF for this case is essentially different than the one shown in
figure \ref{aokiPlot}.

First of all, the PDF seems to be independent of the value of the
external field.  We might appreciate a slight dependence on the
volume, since it seems that the peak height decreases as $V$
increases, but this effect is too small to be significant, and in any
case it is a good approximation to say that the distribution is also
independent of the volume. Comparing this distribution with the former
one (Gibbs state, without external source, figure \ref{aokiPlot}), we
notice that the PDF of the operator with external source is much
wider, and we then expect the spectrum to be different as well.  The
essential difference lies in the low modes of the Dirac operator: in
the case without external source, there is a lower bound given by
$1/V$ for the modulus of any eigenvalue, whereas at $m_t\neq0$ the
eigenvalues can become arbitrarily small \footnote{A consequence of
  this property is the fact that the PDF of $Q$ at $m_t\neq0$ fits
  perfectly to a Lorentzian distribution of infinite tails. One could
  also be concerned that $Q$ become singular if a zero-mode appeared
  in the spectrum. Actually, the set of configurations with a
  zero-mode has measure zero, and $Q$ remains non-singular.}.  Since
the PDF depends strongly on the spectrum of the Dirac operator, we
expect the cases $m_t=0$ and $m_t\neq0$ to be inherently different.

\begin{figure}[h!]
\begin{center}
\resizebox{13 cm}{!}{\includegraphics[angle=270]{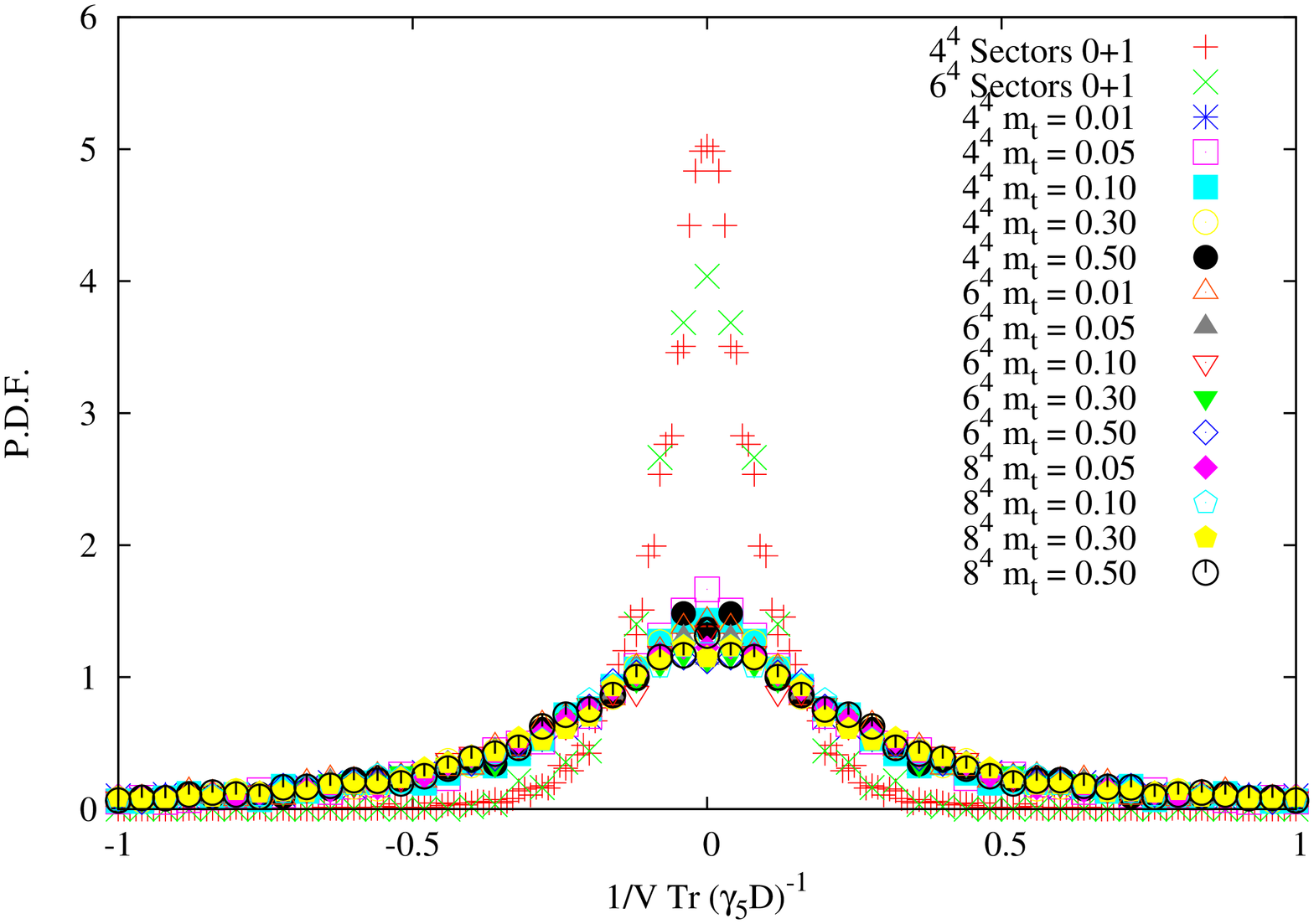}}
\caption{PDF of operator $Q$ inside the Aoki phase for all
  configurations. Comparison of the case $m_t\neq0$ versus the case
  $m_t=0$.\label{TPlot}}
\end{center}
\end{figure}

We think that this result is the strongest indication of additional
vacua structure in the Aoki phase outside the standard picture. Indeed
the results of figure \ref{TPlot} clearly show that the sample of
gauge configurations obtained in the Aoki phase at $m_t = 0$ is
qualitatively different from the sample obtained at $m_t\ne 0$, even
in the $m_t\rightarrow 0$ limit.  On the other hand the differences in
the samples can never come from the additional standard Aoki vacua
since if we change the twisted mass term in the dynamical generation
by any other symmetry breaking term selecting other standard Aoki
vacuum, for instance $i m_t \bar\psi\gamma_5\tau_{1,2}\psi$, the
fermion determinant does not change.

\section{Second moments of the PDFs of fermion bilinears}

Since $P'$ is broken according to our results plus the assumption that 
$Q$ is an intensive operator, the symmetry arguments supporting the
existence of a tower of sum rules are no longer valid, thus it is
natural to explore if the expectation value
$\left\langle\left(i\bar\psi\gamma_5\psi\right)^2\right\rangle$ will
be non-zero in the Gibbs state inside the Aoki phase. This observable
is extremely difficult to measure, nonetheless we managed to obtain
sensible results from our data, which we show in Table \ref{Final}. As
already stated in Section \ref{simulations}, these results assume
continuity of the weights of the different sectors when $m_t
\rightarrow 0$.

\begin{table}
\begin{center}
\footnotesize
\begin{tabular}{|c|c|c|c|c|}
\hline Volume &
$\left\langle\left(i\bar\psi_u\gamma_5\psi_u\right)^2\right\rangle$ &
$\left\langle\left(i\bar\psi\gamma_5\psi\right)^2\right\rangle$ &
$\left\langle\left(i\bar\psi\gamma_5\tau_3\psi\right)^2\right\rangle$ &
$\langle Q^2\rangle$ \\
\hline
$4^4$ Sector 0 & $\left(1.93\pm0.02\right)\times10^{-2}$ & $\left(2.51\pm0.07\right)\times10^{-2}$ & $\left(5.23\pm0.09\right)\times10^{-2}$ & $\left(6.8\pm0.3\right)\times10^{-3}$ \\
$6^4$ Sector 0 & $\left(2.15\pm0.05\right)\times10^{-2}$ & $\left(2.82\pm0.18\right)\times10^{-2}$ & $\left(5.76\pm0.13\right)\times10^{-2}$ & $\left(7.4\pm0.6\right)\times10^{-3}$ \\
$4^4$ Sector 1 & $\left(6.0\pm0.5\right)\times10^{-3}$   & $\left(-5.5\pm0.5\right)\times10^{-2}$  & $\left(7.9\pm0.4\right)\times10^{-2}$   & $\left(3.34\pm0.20\right)\times10^{-2}$ \\
$6^4$ Sector 1 & $\left(1.75\pm0.09\right)\times10^{-2}$ & $\left(0.4\pm0.4\right)\times10^{-2}$   & $\left(6.56\pm0.26\right)\times10^{-2}$ & $\left(1.53\pm0.15\right)\times10^{-2}$  \\
\hline
$4^4$ Total & $\left(1.50\pm0.03\right)\times10^{-2}$ & $\left(-0.6\pm1.9\right)\times10^{-3}$  & $\left(6.08\pm0.19\right)\times10^{-2}$ & $\left(1.54\pm0.08\right)\times10^{-3}$ \\
$6^4$ Total & $\left(1.81\pm0.12\right)\times10^{-2}$ & $\left(1.27\pm0.29\right)\times10^{-2}$ & $\left(6.0\pm0.3\right)\times10^{-2}$   & $\left(1.17\pm0.13\right)\times10^{-3}$ \\
\hline
\end{tabular}
\caption{Interesting v.e.v. for the Aoki phase in the different
  sectors. Only statistical errors are shown.\label{Final}}
\end{center}
\end{table}
\normalsize

The fourth column in this table refers to
$\left\langle\left(i\bar\psi\gamma_5\tau_3\psi\right)^2\right\rangle$,
the landmark of the Aoki phase. As we see, it is clearly non-zero for
all the volumes, confirming that our simulations lie within the Aoki
phase. The second column shows the values for
$\left\langle\left(i\bar\psi_u\gamma_5\psi_u\right)^2\right\rangle$,
which is an order parameter for parity, but it takes into account just
one flavor (which we labeled $u$). This quantity should be non-zero
inside the parity-breaking Aoki phase, regardless of the discussion of
the new vacua. Finally, the most important observable, the flavor
singlet pseudoscalar
$\left\langle\left(i\bar\psi\gamma_5\psi\right)^2\right\rangle$, which
--since we are dealing with two degenerate flavors, $u$ and $d$-- is
the sum of two of the former condensates $i\bar\psi_u\gamma_5\psi_u +
i\bar\psi_d\gamma_5\psi_d$. The standard picture of the Aoki phase
predicts zero expectation value of this parameter in any Aoki vacuum,
whereas each one of the pseudoscalars restricted to one flavor
$i\bar\psi_{u,d}\gamma_5\psi_{u,d}$ will be non-zero due to parity
breaking. Hence, the standard picture of the Aoki phase enforces an
antiferromagnetic ordering of the pseudoscalars
$i\bar\psi_u\gamma_5\psi_u = -i\bar\psi_d\gamma_5\psi_d$, which is not
required in our hypothesis of the new vacua. The data of the third 
column in Table \ref{Final} show a clear non-zero expectation value in
the case of the largest volume $6^4$, supporting our previous
discussion regarding $P'$ breaking.

\begin{table}
\begin{center}
\begin{tabular}{|c|c|c|}
\hline $V$ & $m_t$ & $\left\langle i\bar\psi\gamma_5\tau_3\psi\right\rangle$ \\
\hline
$4^4$ & $0.01$ & $\left(56.00\pm0.06\right)\times10^{-2}$ \\
$4^4$ & $0.05$ & $\left(65.149\pm0.022\right)\times10^{-2}$ \\
$6^4$ & $0.01$ & $\left(53.603\pm0.015\right)\times10^{-2}$ \\
$6^4$ & $0.05$ & $\left(66.888\pm0.008\right)\times10^{-2}$ \\
$8^4$ & $0.05$ & $\left(66.953\pm0.007\right)\times10^{-2}$ \\
$4^4$ & $0.10$ & $\left(75.512\pm0.009\right)\times10^{-2}$ \\
$6^4$ & $0.10$ & $\left(76.291\pm0.004\right)\times10^{-2}$ \\
$8^4$ & $0.10$ & $\left(76.307\pm0.003\right)\times10^{-2}$ \\
$4^4$ & $0.30$ & $\left(92.214\pm0.005\right)\times10^{-2}$ \\
$6^4$ & $0.30$ & $\left(92.309\pm0.002\right)\times10^{-2}$ \\
$8^4$ & $0.30$ & $\left(92.311\pm0.002\right)\times10^{-2}$ \\
$4^4$ & $0.50$ & $\left(96.225\pm0.003\right)\times10^{-2}$ \\
$6^4$ & $0.50$ & $\left(96.241\pm0.001\right)\times10^{-2}$ \\
$8^4$ & $0.50$ & $\left(96.243\pm0.001\right)\times10^{-2}$ \\
\hline
\end{tabular}
\caption{Evolution of the v.e.v. of the Aoki parameter as a function
  of the twisted mass external field.\label{AokiTmTab}}
\end{center}
\end{table}

\begin{figure}[h!]
\begin{center}
\resizebox{13 cm}{!}{\includegraphics[angle=270]{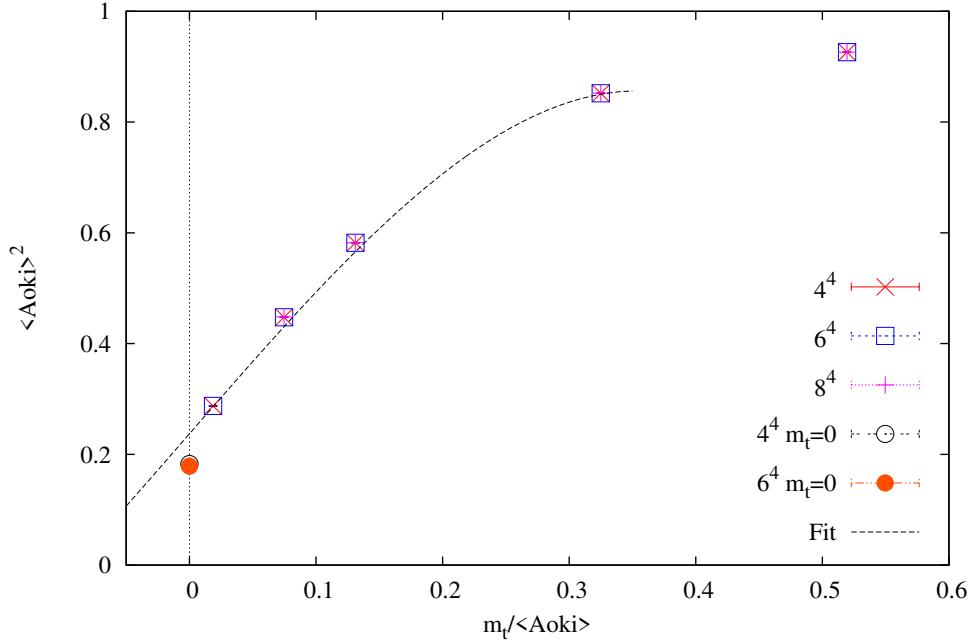}}
\caption{Fisher plot of the evolution of the v.e.v. of the Aoki
  parameter as a function of the twisted mass external field. The
  circles show the prediction obtained from the simulations in the
  Gibbs state (see eq.\eqref{standard}). The fit shown is meant to
  guide the eye, and shouldn't be taken as a prediction of the
  behavior of the v.e.v. of the Aoki parameter. The fit should be more
  reliable for small values of $m_t$. \label{AokiTmPlot}}
\end{center}
\end{figure}

Concerning equation \eqref{standard}, which relates the mean value of
the square Aoki condensate in the Gibbs state ($\epsilon$-regime) with
the expectation value of this order parameter in the vacuum selected
by a twisted-mass term ($p$-regime), and since we have data for the
Aoki condensate at several lattice sizes and twisted masses (see Table
\ref{AokiTmTab}), we can check its plausibility. This is a relevant
test since, as discussed in Section 2.3, equation \eqref{standard}
should be realized in the standard Aoki scenario
\cite{pdc,sharpe2}. To this end we report in figure \ref{AokiTmPlot}
our results for the Aoki condensate at several values of the
twisted-mass, $m_t$, in $4^4, 6^4$ and $8^4$ lattices. The circles in
the ordinates axis stand for the values of the Aoki condensate at $m_t
= 0$ in $4^4$ and $6^4$ lattices, obtained from equation
\eqref{standard}, and using as input our results for
$\left\langle(i\bar\psi\gamma_5\tau_3\bar\psi)^2\right\rangle$
reported in Table \ref{Final}. The figure, which is a Fisher plot
\cite{fisher}, shows that any reliable extrapolation of the data to
$m_t = 0$ gives a value for the Aoki condensate larger than the one
obtained from equation \eqref{standard}.

One could argue that, at the volumes we are working, the finite size
effects should be large and noticeable, and that the inclusion of
these effects could lead to an agreement between our measurements at
$m_t=0$ and $m_t\neq0$. Nonetheless, the data at $m_t=0$ --which we
expect to suffer more these finite volume effects due to the existence
of massless pions-- reveals that these effects are not very large, for
the value of the Aoki condensate remains very stable after a fivefold
increase in the volume. 

\section{Relation with $W\chi PT$}

The results reported in this paper on the vacuum structure of the Aoki
phase do not agree, as follows from the previous sections, with the
predictions of $W\chi PT$ of Sharpe and Singleton \cite{sharpe} for
this phase. Since $W\chi PT$ has been successfully applied in many
contexts, it is worthwhile to analyze the possible origins of this
discrepancy.

First one should notice that the chiral effective Lagrangian is based
on the continuum effective Lagrangian written as a series of
contributions proportional to powers of the lattice spacing $a$, plus
the construction of the corresponding chiral effective Lagrangian,
keeping only the terms up to order $a^2$ \cite{sharpe}. This means
that predictions based on the use of this chiral effective Lagrangian
should work close enough to the continuum limit, where keeping terms
up to order $a^2$ can be justified. However our investigation of the
Aoki phase has been done at $\beta=2.0$, and a very rough estimation
gives a lattice spacing at this $\beta$ of order $3.0 \;
\text{GeV}^{-1}$. Hence a possible explanation for the discrepancies
found relies on the necessity of including higher order terms in the
chiral effective Lagrangian.

The second point to notice is that our simulations were performed deep
in the Aoki phase ($\kappa = 0.25$) and hence far away from the
critical line where the neutral pion is massless. The chiral effective
Lagrangian approach is based on the assumption that the relevant low
energy degrees of freedom are the three pions. This assumption is very
reliable in the physical phase near the critical line, and also in the
Aoki phase near the critical line, but it could break down as we go
deep in the Aoki phase, as is our case. To better understand this
point we have analyzed the two flavor Nambu-Jona Lasinio model with
Wilson fermions in the mean field approximation not only with the
standard action \cite{bitar, aokinjl} but with the more general action
with an $SU(2)_V\times SU(2)_A\times U(1)_B$ symmetry. The continuum
Euclidean Lagrangian density for this model is
\be
{\cal L} = - \bar\psi\left(\slashed{\partial} + m \right) \psi +
            G_1 \left[ \left(\bar\psi\psi\right)^2 + \left(i \bar\psi
            \gamma_5{\vec{\tau}}  \psi\right)^2 \right] +
            G_2 \left[ \left(i \bar\psi \gamma_5\psi\right)^2 + \left(\bar\psi
            {\vec{\tau}}  \psi\right)^2 \right]
\label{NJL}
\ee
This model regularized on a hypercubic four-dimensional lattice with
Wilson fermions was analyzed at $G_2=0$ in the mean field or first
order $1/N$ expansion by Bitar and Vranas \cite{bitar} and Aoki et
al. \cite{aokinjl}, who found a phase, for large values of $G_1$, in
which both, flavor symmetry and parity, are spontaneously broken, in
close analogy to lattice QCD with Wilson fermions.

The Nambu-Jona Lasinio model given by action \eqref{NJL} enjoys the
same $SU(2)_V\times SU(2)_A\times U(1)_B$ symmetry of $QCD$ and 
it is an effective model
to describe the low energy physics of $QCD$ \cite{shankar}. As stated
before we have analyzed the phase diagram of this model in the mean
field approach in the three parameters space ($\kappa$, $G_1$, $G_2$),
and a detailed report of our results will appear in a forthcoming
publication. What we want to point out here is that whereas in the
large $G_1$ and small $G_2$ region the results on the phase diagram
agree with those obtained in \cite{aokinjl} and therefore with the
standard Aoki picture, at larger values of $G_2$ we find phases where
the flavor singlet pseudoscalar condensate $i\bar\psi\gamma_5\psi$
takes a non-vanishing vacuum expectation value, including phases with
vacua degenerated with the standard Aoki vacua, and which can not be
connected with the Aoki vacua by parity-flavor symmetry
transformations.  This example shows that the complete phase diagram
of a model having the same chiral, vector and discrete symmetries as
$QCD$, cannot be understood assuming that chiral symmetry is realized
in the standard strong interaction Goldstone picture.

\section{The condensates near the $m_0 = -4.0$ line of the phase diagram}

The phase diagram of lattice QCD with two degenerate flavors of Wilson
fermions in the bare fermion mas, $m_0$, and gauge coupling, $g^2$, plane is
symmetric under the $m_0\rightarrow -(m_0 + 8)$ change. At the invariant point,
$m_0 = -4$, the system shows extra symmetries. Indeed, if we parameterize the
bare fermion mass as $m_0 = -4 + \epsilon$, the fermionic Wilson action for one
flavor can be written as
\begin{multline}
{S_F} = \epsilon \sum_x \bar\psi(x)\psi(x) - {1\over 2} \sum_{x, \mu} 
\left(\bar\psi(x) (1 + \gamma_\mu) U^+_\mu (x - \mu)\psi(x-\mu) +\right. \\ 
\left.\bar\psi(x) (1 - \gamma_\mu) U_\mu (x + \mu)\psi(x+\mu)\right) 
\label{wact}
\end{multline}
and at $\epsilon = 0$ ($m_0 = -4$) we have an extra global $Z(4)$ symmetry
for each flavor, generated by the following transformations
$$
\psi'(x) = (-1)^{x_1+x_2+x_3+x_4} i \psi(x)
$$
\be
\bar\psi'(x) = \bar\psi(x) (-1)^{x_1+x_2+x_3+x_4} i
\label{symm}
\ee
These symmetries enforce, in the two flavor model, the equality of the
condensates  
\be
\langle \left(i\bar\psi\gamma_5\tau_3\psi\right)^2 \rangle = 
\langle \left(i\bar\psi\gamma_5\psi\right)^2 \rangle
\label{condensados}
\ee
and since the standard wisdom on the phase diagram of this model, corroborated
on the other hand by the strong coupling and large number of colors expansion
of \cite{a1,a2}, tells us that $\epsilon = 0$ is in the Aoki phase 
for any value of the
gauge coupling, we should conclude that there exists at least a line in the
phase diagram, $\epsilon = 0$, along which both condensates take a non-vanishing
identical value. Notice also that these non-vanishing condensates imply the
spontaneous breaking of not only parity and flavor symmetries, but also the
$Z(4)$ symmetries \eqref{symm}.

On a finite lattice, the two condensates \eqref{condensados} can be described by
the ratio of two even polynomials of $\epsilon$, and a simple Taylor expansion
would suggest that the two condensates do not vanish in a region of non-vanishing
measure of the phase diagram, thus giving a qualitative picture consistent with
the numerical results reported in this article. However the actual situation can
be more complex, since we can not exclude a priori that the convergence radius of
the $\epsilon$-expansion vanishes in the infinite volume limit. This convergence
radius is given by the distance to the origin of the nearest zero of the partition
function in the complex $\epsilon$-plane, and the scaling of these zeroes with the
lattice volume is related to the chiral condensate $\bar\psi\psi$, in a similar way
as the zeroes of the partition function of the staggered formulation in the
complex mass plane are related to the realization of the chiral symmetry for
staggered fermions. This analogy comes from the fact that at $\epsilon=0$ the
chiral condensate in the Wilson formulation vanishes due to the $Z(4)$ symmetry,
in the same way as the chiral condensate in the staggered formulation vanishes at
zero fermion mass because of chiral symmetry. Also, the behavior of the chiral
condensate when $\epsilon\rightarrow 0$, is controlled by the scaling of the zeroes
of the partition function in the complex $\epsilon$ plane with the lattice volume.
A singular condensate at $\epsilon=0$ will be obtained if the zeroes of the
partition function approach the point $\epsilon=0$ with the lattice volume. On the
contrary, if the scaling of the zeroes stops at finite distance, an analytical
value for the chiral condensate will be obtained.  

The solution of this problem outside approximations is a very hard task. However
the dependence of the chiral condensate on $\epsilon$ was computed by Aoki in the
strong coupling and large N expansion in \cite{a1,a2}, the final result being 
\be
{{\langle \bar\psi\psi \rangle}\over{8N}} = {{3\epsilon}\over{16 - \epsilon^2}}
\label{chco}
\ee
for $0\le\epsilon^2<4$. Equation \eqref{chco} shows that the chiral condensate
is an analytical function of $\epsilon$ in the strong coupling limit, and that
the convergence radio of the $\epsilon$-power expansion is 2.
Hence we should expect the same analyticity domain for the other two 
condensates \eqref{condensados}.

These results show that, at least in the strong coupling and large N
approximation, we should expect a phase structure for QCD with two flavors of
Wilson fermions as the one proposed by us, giving theoretical support to the
numerical results reported in this article.

\section{Conclusions and Outlook}

Three years ago the standard scenario for the phase structure of
lattice QCD with two degenerate flavors of Wilson fermions was revised
by three of us in \cite{monos}, where we conjectured on the appearance
of new vacua in the Aoki phase, which can be characterized by a
non-vanishing vacuum expectation value of
$i\bar\psi_u\gamma_5\psi_u+i\bar\psi_d\gamma_5\psi_d$, and which can
not be connected with the Aoki vacua by parity-flavor symmetry
transformations. However, Sharpe pointed out in \cite{sharpe2} that
the standard picture for the Aoki phase can be understood using the
chiral effective theory appropriate to the Symanzik effective action,
and that within this standard analysis, the flavor singlet
pseudoscalar expectation value vanishes, $\left\langle
i\bar\psi\gamma_5\psi\right\rangle = 0$. As the standard scenario for
the Aoki phase is indeed a direct consequence of the $W\chi PT$
application to this phase, we were also calling into question in
\cite{monos} the validity of the $W\chi PT$ analysis, at least for low
values of $\beta$, and therefore large values of the lattice spacing
$a$. These issues are relevant enough to make it worthwhile to
continue these investigations in order to clarify the actual scenario
for the Aoki phase.

For the last few years we have performed an extensive research on the
vacuum structure of the Aoki phase, in order to find out if our
alternative vacuum structure, derived from a PDF analysis, was
realized or not. These simulations, which have been mainly performed
in the absence of any parity-flavor symmetry breaking external source,
are plagued by technical difficulties which have been responsible for
the slow progress in the field. Indeed the addition of a twisted mass
external source, as it has been done in the past simulations of the
Aoki phase, automatically selects the vacuum where the standard
properties of the Aoki phase are verified. This point could explain
why nobody ever saw a new phase like the one we are proposing, since
all the past simulations performed with two flavors of Wilson fermions
inside the Aoki phase were done within an external twisted mass term.

Notwithstanding these difficulties, we have provided in this work
evidence pointing to a more complex vacuum structure in the Aoki phase
of two flavor QCD, as conjectured in \cite{monos}. Indeed the results
reported in Table \ref{Final}, which were obtained under the
assumption that the weights of the different sectors are continuous at
$m_t = 0$, show how we were able to perform a direct measurement of
$\left\langle\left(i\bar\psi\gamma_5\psi\right)^2\right\rangle$ in the
Aoki phase, which gave a non-zero value for this operator in the $6^4$
lattice, of the same order of magnitude as
$\left\langle\left(i\bar\psi_u\gamma_5\psi_u\right)^2\right\rangle$,
and
$\left\langle\left(i\bar\psi\gamma_5\tau_3\psi\right)^2\right\rangle$,
the last two being non-vanishing in the Aoki phase in the standard
scenario.  Thus this result points to the breaking of the hypothesis
of the sum-rules \eqref{rela}. Furthermore a check of equation
\eqref{standard}, a equation which should hold if the standard
scenario for the Aoki phase is realized, reported in figure 8, points
out to a more complex vacuum structure too.

However the strongest indication, in our opinion, on the existence of
a vacuum structure in the Aoki phase, more complex than the one of the
standard picture, comes from our analysis of the PDF of the operator
$Q$ \eqref{traza2}. Our motivation for the analysis of this kind of
density of "topological charge" operator, which measures the
asymmetries in the eigenvalue distribution of the Hermitian
Dirac--Wilson operator, was twofold.  First it appears in the
computation of the second moment of the PDF of the flavor singlet
pseudoscalar order parameter \eqref{vasp}, and second $Q$ is an order
parameter for the symmetry $P'$, composition of parity and discrete
flavor transformations, described in Section 2.

Our numerical results for the PDF of $Q$, together with the assumption
that $Q$ is an intensive operator, suggest that the $P'$ symmetry is
realized in the vacuum of the physical phase, but not in the Aoki
phase, at least in the strong coupling region we have analyzed. 
However the cleanest signal of further structure in the Aoki phase
comes from the results on the PDF of $Q$ reported in figure
\ref{TPlot}. Those results clearly show that the sample of gauge
configurations obtained in the Aoki phase at $m_t = 0$ is
qualitatively different from the sample obtained at $m_t\ne 0$, since
they give incompatible PDF's.  Furthermore this result stays true in
the $m_t\rightarrow 0$ limit, as follows from the independence of the
shape of the PDF of $Q$ on the twisted mass $m_t$.  On the other hand
the differences in the samples can never come from the additional
standard Aoki vacua since if we change the twisted mass term in the
dynamical generation by any other symmetry breaking term selecting
other standard Aoki vacuum, as for instance $i m_t
\bar\psi\gamma_5\tau_{1,2}\psi$, the fermion determinant does not
change.

As the sum rules \eqref{rela} required in the standard scenario are
only supported, as discussed before, by the $W\chi PT$ analysis of the
Aoki phase, our results could call into question also the validity of the
$W\chi PT$ analysis performed in \cite{sharpe} for low values of
$\beta$ (around 2.0, very coarse lattices) and deep into the Aoki
phase ($\kappa = 0.25$). However, $W\chi PT$ is expected to work at
higher values of $\beta$, near the continuum limit and close to the
critical line. Indeed, one could be concerned that, at such a low
$\beta$ and high $\kappa$, we are far from the continuum limit and
from the region of applicability of the standard chiral effective
Lagrangian, as discussed in Section 6. Actually the analysis done 
in Section 7 gives theoretical support to our alternative scenario. 
Nonetheless our work is devoted to the Aoki phase on the lattice, which
might not even have a continuum limit. 

In any case any improvement of our results in larger lattices and for
more $\beta, \kappa$ values would be of great interest. Unfortunately, 
given the technical difficulties of the simulations inside the Aoki
phase, this calculation is outside our present computing resources.

\section*{Acknowledgments}

It is a pleasure to thank Steve Sharpe for his invaluable help,
interesting comments and fruitful discussions; his remarks allowed us
to improve greatly this work. We also want to thank Fabrizio Palumbo
for useful discussions. This work was funded by an INFN-MICINN
collaboration (under grant AIC-D-2011-0663), MICINN (under grants
FPA2009-09638 and FPA2008-10732), DGIID-DGA (grant 2007-E24/2) and by
the EU under ITN-STRONGnet (PITN-GA-2009-238353). E. Follana is
supported on the MICINN Ram\'on y Cajal program, and A. Vaquero is
supported by the Research Promotion Foundation (RPF) of Cyprus under
grant $\Pi$PO$\Sigma$E$\Lambda$KY$\Sigma$H/NEO$\Sigma$/0609/16.

\end{document}